\documentclass[12pt]{article}

\def\rdots{\mathinner{\mkern1mu\raise1pt\vbox{\kern1pt\hbox{.}}\mkern2mu
   \raise4pt\hbox{.}\mkern2mu\raise7pt\hbox{.}\mkern1mu}}
\newcommand{\Z}{{\rm Z\kern-.35em Z}}
\newcommand{\bP}{{\rm I\kern-.15em P}}
\newcommand{\Q}{\kern.3em\rule{.07em}{.65em}\kern-.3em{\rm Q}}
\newcommand{\R}{{\rm I\kern-.15em R}}
\newcommand{\h}{{\rm I\kern-.15em H}}
\newcommand{\C}{\kern.3em\rule{.07em}{.65em}\kern-.3em{\rm C}}
\newcommand{\T}{{\rm T\kern-.35em T}}

\newcommand{\be}{\begin{equation}}
\newcommand{\ee}{\end{equation}}

\newcommand{\al}{\alpha}

\begin{document}
  
\openup 1.5\jot
\centerline{e to the A, in a New Way, Some More to Say}

\vspace{1in}
\centerline{Paul Federbush}
\centerline{Department of Mathematics}
\centerline{University of Michigan}
\centerline{Ann Arbor, MI 48109-1109}
\centerline{(pfed@math.lsa.umich.edu)}

\vspace{1in}

\centerline{\underline{Abstract}}

Expressions are given for the exponential of a hermitian matrix, $A$.
Replacing $A$ by $iA$ these are explicit formulas for the Fourier
transform of $e^{iA}$.  They extend to any size $A$ the previous results
for the $2 \times 2, \ 3 \times 3$, and $4 \times 4$ cases.  The
expressions are elegant and should prove useful.
\vfill\eject

The support of the Fourier transform of $e^{iA}$ was established by E.
Nelson in [1].  (That is the Fourier transform of each entry of the matrix
$e^{iA}$ in terms of the entries of the hermitian matrix $A$.)  But I
believe this result of E. Nelson is very little known in the mathematical
community at large.  In further work [2], [3], [4] the transform was
exhibited in the $2 \times 2$ case, and presented in some unwieldy forms
in higher dimensions. In a previous paper,[5], explicit formulas were
obtained  for $2\times2, 3 \times 3$ and $4 \times 4$ matrices.  We here
treat the general case.

Let $A$ be an $r \times r$ hermitian matrix.  We write
\be	{\rm Det}(1-A) = \sum^r_{j=0} P_j(A)	\ee
where $P_j(A)$ is homogeneous of degree $j$ in the entries of $A$.  The
formulas we obtain for the exponential of $A$ are as follows:
\be
(e^A)_{\al\beta} = \frac 1{\Gamma(r)} \sum^r_{j=0} P_j(A) \frac
{d^{r-j}}{ds^{r-j}} \left( \int d\Omega e^{sTr(AW)} W_{\al\beta}
s^r\right) \bigg|_{s=1}
\ee
or 
\be
(e^A)_{\al\beta} = \frac 1{\Gamma(r)} \sum^r_{j=0} P_j(A) \frac
{d^{r-j}}{ds^{r-j}} \left( \int d\Omega e^{s<A\vec{n},\vec{\bar{n}}>}
n_\al\bar{n}_\beta s^r\right) \bigg|_{s=1}
\ee
Here $\vec n$ is a unit vector in $\C^r$, and $W_{ij} = n_i \bar n_j$, a
rank one hermitian matrix.  $\int d\Omega$ denotes a normalized integral
over all such $\vec n$, an integral over the unit sphere in $\C^r$ with
unitary-invariant measure.  That the support of the Fourier transform lies
on the complex projective space of such $W$ is the content of Nelson's
theorem.

In fact the formulas in (2) and (3) do not coincide with formulas in [5]
when $r=2,3$ or 4, but formulas {\it of such type} are not unique.  We do
not know the full scope of such non-uniqueness.

We first sketch a derivation/proof of formulas (2) and (3), especially
emphasizing the {\bf ideas}.  We note the relation between gaussian
integrals in $d=2r$ real dimensions, and integrals over the corresponding
unit sphere $S^{d-1}$.
\begin{eqnarray}
 \frac 1 {\cal N} \int dx_ie^{-\Sigma x^2_i} \prod^{2N} x_{\al(i)} 
&=& \frac 1 {\cal N} \int r^{d-1} r^{2N} e^{-r^2} dr\int d\Omega'
\prod^{2N} n_{\al(i)} \\
&=& \int_{S^{d-1}} d\Omega \prod^{2N}n_{\al(i)} \cdot \frac{\int
dre^{-r^2}r^{2N+d-1}}{\int dr e^{-r^2} r^{d-1}} \\
&=& \int_{ S^{d-1}} d\Omega \prod^{2N}n_{\al(i)}
\frac{\Gamma\left(\frac{2N+d}{2}\right)} {\Gamma\left( \frac {d}{2}
\right)} .
\end{eqnarray}
Here $n_i$ is the unit vector parallel to $x_i$, and $\int d\Omega'$ is
integral over the sphere in its usual measure and $\int d \Omega$ the
normalized spherical measure.  From (6) we see {\it the integral over a
unit sphere of a homogeneous polynomial of degree $2N$ is ``approximately"
$1/N!$  the gaussian integral of the same polynomial.}

We note that multiplying the term in $A^k$ by $\frac 1 {k!}$ induces a
transform (formally) as follows
\be	1 + A + A^2 + \cdots = \frac 1{1-A} \longrightarrow e^A \ .
\ee

We consider the gaussian integral formula (for $|A| < 1$):
\be
\frac{{\rm Det}(1-A)}{{\cal N}} \int dx_i
e^{-\Sigma|x^2_i|+<A\vec{x},\vec{\bar{x}}>} x_\al\bar{x}_\beta = \left(
\frac 1{1-A}\right)_{\al \beta}.
\ee

In the expansion of the integrand on the left side of (8) each power of
$A$ has associated to it two powers of $x$.  Thus converting from a
gaussian integral to an integral over a unit sphere approximately
multiplies each power of $A^N$ by $\frac 1 {N!}$, which would convert
$\frac 1 {1-A}$ to $e^A$.  The {\bf wrong} formula we get putting these
ideas together would yield:
\be
`` {\rm Det}(1-A) \int d\Omega e^{<A\vec{n},\vec{\bar{n}}>}
n_\al\bar{n}_\beta = (e^A)_{\alpha\beta} "
\ee

We turn to the easy task of converting the above careless argument leading
to the wrong formula (9), to the detailed correct computation that turns
(9) into (2).  (Hitherto we have trodden a path redolent with the creative
epiphanies of mathematical research, we now segue to the ineluctable
concomitant consecration to inferential syntax.)   We start from the right
side of equation (3), and assume for the moment $|A| < 1$.
\be
 \frac 1{\Gamma(r)} \sum^r_{j=0} P_j(A) \frac {d^{r-j}}{ds^{r-j}} \left(
\int d\Omega e^{s<A\vec{n},\vec{\bar{n}}>} n_\al\bar{n}_\beta s^r\right)
\bigg|_{s=1}
\ee
We expand the exponent and perform the operations on $s$, getting
\be
\sum^r_{j=0} P_j(A) \int d \Omega \sum^\infty_{k=0} \Big(
<A\vec{n},\vec{\bar{n}}>\Big)^k n_\al \bar{n}_\beta \cdot 
\frac 1{\Gamma(r)} \cdot \frac 1{k!} \cdot \frac {(r+k)!}{(k+j)!}
\ee
Now we use the equality of equations (4), (5), (6) to convert (11) to
\be
\sum^r_{j=0} P_j(A)\frac 1 {\cal N} \int dx_i e^{-\Sigma x^2_i}
\sum^\infty_{k=0} \Big( <A\vec{x},\vec{\bar{x}}>\Big)^k x_\al
\bar{x}_\beta \frac 1{\Gamma(r)}  \frac 1{k!}  \frac {(r+k)!}{(k+j)!}
\frac {\Gamma(\frac d 2)}{\Gamma(\frac{2(k+1)+d}{2})}
\ee
We rewrite the equality of equation (8) in expanded form
\be
\sum^r_{j=0} P_j(A)\frac 1 {\cal N} \int dx_i e^{-\Sigma x^2_i}
\sum^\infty_{k=0} \Big( <A\vec{x},\vec{\bar{x}}>\Big)^k x_\al
\bar{x}_\beta \frac 1{k!} = 1 + A + A^2 + \cdots
\ee
We mark the fact that equation (13) is a separate equality for each
homogeneous degree in powers of $A$.
On each side of the equation we multiply terms homogeneous of degree
$\ell$ by $\frac 1{\ell !}$, arriving at
\be
\sum^r_{j=0} P_j(A)\frac 1 {\cal N} \int dx_i e^{-\Sigma x^2_i}
\sum^\infty_{k=0} \Big( <A\vec{x},\vec{\bar{x}}>\Big)^k x_\al
\bar{x}_\beta \frac 1{k!} \frac 1{(j+k)!}= 1 + A + \frac{A^2}{2!} + \cdots
= e^A .
\ee
The equality of the left side of (14) with (12) follows from
\be
\frac 1{\Gamma (r)}  \frac 1{k!} \cdot  \frac {(r+k)!}{(k+j)!} \cdot
\frac {\Gamma(\frac d 2)} {\Gamma(\frac{2(k+1)+d}{2})} = \frac 1{k!} \frac
1{(j+k)!}
\ee
using $r=2d$.

We have thus established our equalities of equations (2) and (3) for $|A|
< 1$.  But each side of these equations is analytic in the elements of
$A$, so the equalities hold for all hermitian $A$.

\underline{Acknowledgment}:  I would like to thank Alexander Barvinok for
an all important discussion on evaluating integrals over the unit sphere.

\vfill\eject

\centerline{References}

\begin{itemize}
\item[[1]]  E. Nelson, {\it Operants: A functional calculus for
non-commuting operators}, Functional Analysis and Related Fields,
Proceedings of a conference in honor of Professor Marshal Stone (Univ. of
Chicago, May 1968) (F.E. Browder, ed.), Springer-Verlag, Berlin,
Heidelberg, and New York, 1970, pp. 172-187.  MR 54:978.
\item[[2]]  B. Jefferies, ``The Weyl Calculus for Hermitian Matrices",
{\it Proc. A.M.S.} {\bf 124} (96) p. 121-128.
\item[[3]] M.E. Taylor ``Functions of Several Self-Adjoint Operators",
{\it Proc. A.M.S.} {\bf 19} (1968), 91-98.  MR {\bf 36}:3149.
\item[[4]] R.F.V. Anderson, ``The Weyl Functional Calculus", {\it J. Func.
Anal.} {\bf 4} (1969) 240-267.  MR {\bf 58}:30405.
\item[[5]] P. Federbush, ``e to the A, in a New Way", math-ph/9903006, to
be published in the {\it Michigan Math. Journal}.

\end{itemize}
\end{document}